\documentclass[11pt]{article}

\usepackage{float}
\usepackage{graphicx,color}
\usepackage{epsfig}
\usepackage{amsmath,amssymb,amsthm}
\usepackage{subfigure}
\usepackage{dsfont}
\usepackage{psfrag}
\usepackage[ruled,noline]{algorithm2e}

\def\1{\mathbf{1}}

\graphicspath{{figures/}}

\floatstyle{ruled}
\newfloat{Algorithm}{tbp}{lop}[section]

\begin{document}

\title{Mapping and Reducing the Brain on the Cloud}

\author{Esha Sahai and Tuhin Sahai}


\maketitle
\begin{abstract}
The emergence of cloud computing has enabled an incredible growth in available hardware resources at very low costs. These resources are being increasingly utilized by corporations for scalable analysis of ``big data'' problems. In this work, we explore the possibility of using commodity hardware such as Amazon EC2 for performing large scale scientific computation. In particular, we simulate interconnected cortical neurons using MapReduce. We build and model a network of $1000$ spiking cortical neurons in Hadoop, an opensource implementation of MapReduce, and present results.
\end{abstract}

\section{Introduction}

Cloud computing is a term that has recently received significant buzz and interest. It has been popularized by social media outlets as well as large companies that provide IT services. Though the meaning of the term ``cloud computing" is context driven, typical usage refers to  software or computer program execution on commodity servers at a location unknown to the user~\cite{cit:IntroCloud}. More importantly, the implementation of underlying cloud computing software (such as Hadoop) allows one to exploit distributed computation paradigms to obtain scalable solutions with fault tolerance~\cite{hadoop}. Cloud computing has been credited with providing scalable infrastructure to Google.~\cite{Dean_h}, Facebook and Twitter to name a few.

The vast trove of user generated data on the Internet has also propelled the emergence of cloud computing infrastructure~\cite{AmazonEC2}. Besides the immediate impact of powering Internet services, cloud computing is slowly impacting other areas of applications, whose ramifications may one day be much greater than simply organizing information on the Internet. It has converted the accessibility to large scale parallel computers from an exclusive club of rich universities and institutions to a commodity that can be purchased via the web. For example, Amazon's Elastic Compute Cloud (EC2 in short) allows users to rent a computer processor for as low as $\$0.02$ an hour~\cite{AmazonEC2}. Thus, for a few hundred dollars one can rent an army of computers to solve any computer problem at hand.

In this work, we investigate the possibility of using cloud services for scientific computation. Applications involving scientific computation can easily require the immense computing resources of supercomputers to extract an answer with required accuracy. For example, IBM researchers simulated a portion of the mouse brain (comprising of $22$ million neurons) on a BlueGene L supercomputer with $8192$ processors~\cite{IBM_Brain}.

\begin{figure}[htb]
\begin{center}
\includegraphics[width=4in]{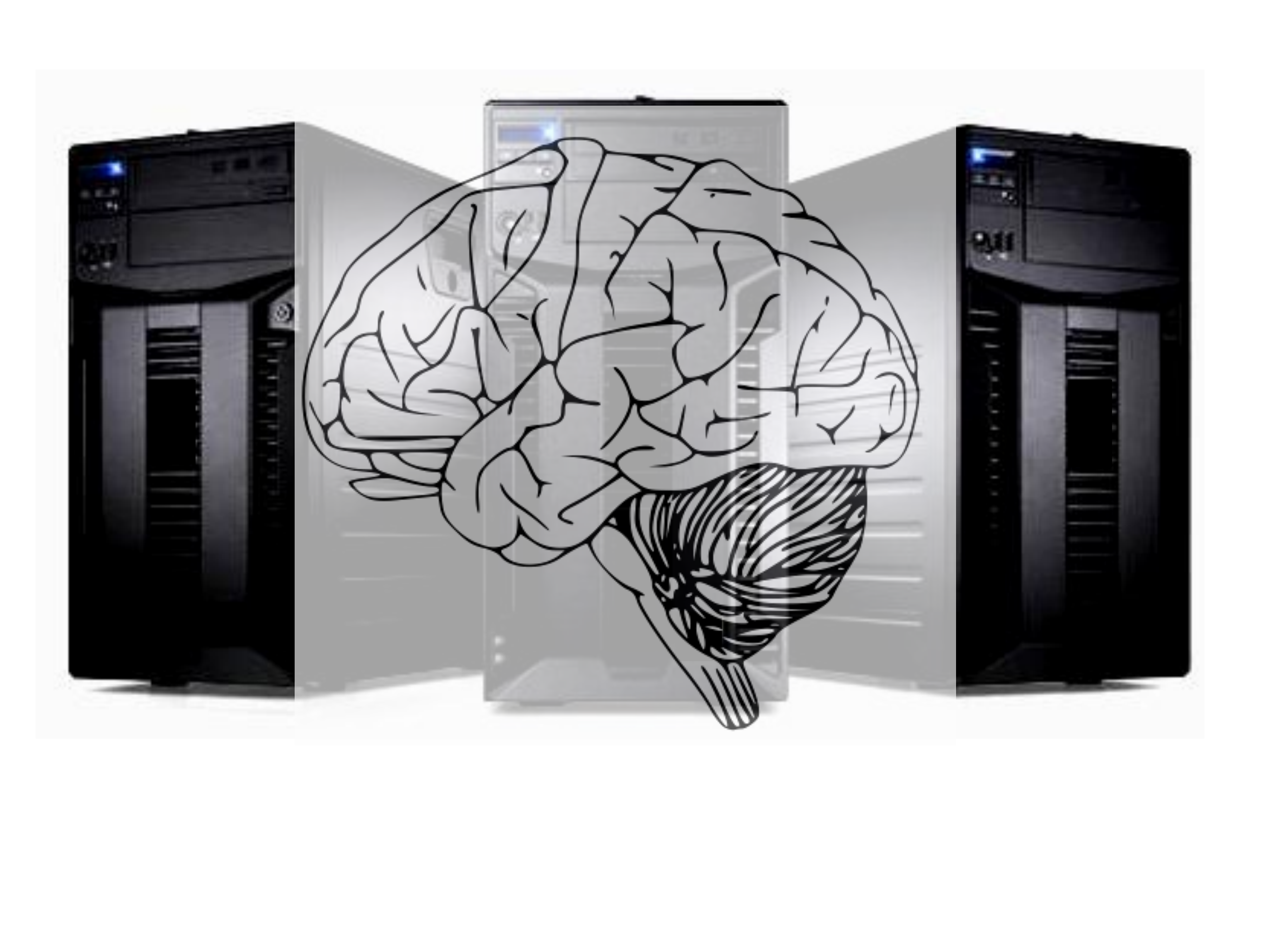}
\caption{We investigate the possibility of using cloud infrastructure for scientific computation. In particular, we use a model of interconnected neurons.}
\label{Fig:cloudbrain}
\end{center}
\end{figure}

 As an illustration of the importance of scientific computation, Oak Ridge National Laboratory has built the Jaguar supercomputer with $299,008$ processors. Parallel computers are routinely used to model the climate~\cite{climate}, design aircraft~\cite{aircraft} and study nuclear fusion~\cite{fusion} to name a few. The available computer time for Jaguar is allocated using a competitive proposal process. Due to the utility of parallel machines, scientific institutions, universities and corporations typically invest millions of dollars annually to build or buy compute time on supercomputers.

 In this manuscript, we investigate if cloud computing is a feasible approach for conducting scientific computations. If it is, then one can simply buy computing time on the cloud as a pay-as-you-go service. This eliminates the need for maintaining large and expensive computing facilities. Note that cloud computing has previously been used to analyze the human genome~\cite{Gene_MR}.

\section{Brief Overview of MapReduce and Hadoop}
As mentioned previously, the cloud computing paradigm has enabled orders of magnitude improvement in the ability to cheaply process large amounts of data~\cite{Dean_h, AmazonEC2}. Arguably, the most popular cloud computing approach that enables scalability is MapReduce~\cite{hadoop, Dean_h}.

A MapReduce task is typically broken down into two
steps: a map step which is followed by a reduce step; thus giving this parallel computing paradigm the name ``MapReduce''. Note that certain applications may only require map steps or alternatively necessitate multiple reduce steps. The input data for a MapReduce job is first fed into a user specified map function.
The map function processes this data and outputs multiple
key-value pairs.  Note that the keys in these key-value pairs need not be unique. For example, in the word count example, keys are words that may be repeated several times in a document.  After the map step completes data processing, the values with the same key are combined locally at every compute node using a combine step~\cite{MapRedbook}. This combine step can be viewed as a local reduce step that minimizes the amount of data transfer between nodes required during the final reduce step.  The key and value list from the combine step is then input to a user-defined reduce step.  The reducer performs computation on this list of keys and values to output a final key-value pair that contains the answer. In case of iterative and graph algorithms, one may need to chain a sequence of map and reduce steps~\cite{MapRedbook}.

Despite criticism~\cite{HadoopDetract},
MapReduce remains a popular cloud computing paradigm that allows user-defined parallel processing of large data sets. MapReduce has been used to implement a host of different algorithms~\cite{MapRedbook}. These include,
word count~\cite{MapRedbook}, Page-rank
algorithm~\cite{PageRank-Brin-Mot}, singular value decomposition~\cite{Mahout} and clustering~\cite{Mahout} to name a few. Once an algorithm is cast into the MapReduce paradigm~\cite{Dean_h}, it becomes
parallelized and ready for execution on large clusters of computers~\cite{hadoop}. Multiple map
jobs are executed in parallel to enable parallelization of input data processing. Similarly, independent reduce jobs are also executed in parallel, enabling scalability of data processing.

The most popular MapReduce implementation is
Hadoop~\cite{hadoop}. Hadoop includes the Hadoop Distributed
File System (HDFS) that manages data. It also automatically schedules jobs, restarts processes in case of machine failures, and handles data transfer between computers. Hence, once an algorithm is cast in the high-level MapReduce setting, Hadoop handles the details of the implementation of the parallel computation.

In the Hadoop system~\cite{hadoop}, NameNode was, until recently, a single point of failure. This critical compute node performs data and job assignments for nodes in the cluster and tracks job failure. This drawback has since been overcome using HA NameNode.

There are several benefits of
MapReduce. The primary benefit is that MapReduce has been implemented in multiple open-source projects including Hadoop. This provides the user significant support with several examples available for demonstration and testing. We thus choose Hadoop based MapReduce to implement our neural models on the cloud.

\section{Neural Models}
Computational neuroscientists routinely use computer models to investigate brain dynamics~\cite{Izhikevich-book,Brain-chaos}. In particular, they have used computational models to explain various phenomena such as infant visual foraging~\cite{Infantforaging}, synchrony~\cite{synchrony} and movement~\cite{movement}.

Brain models are typically constructed by interconnecting models of individual neurons~\cite{IBM_Brain,Brain-net-silva}. The Hodgkin-Huxley equations are an extremely popular model for describing the behavior of individual neurons~\cite{Hodgkin-Huxley}. In this work, we instead use the simple spiking model of a neuron developed in~\cite{Izhikevich}. This model retains the accuracy of the Hodgkin-Huxley model~\cite{Hodgkin-Huxley}, without making unrealistic simplifications routinely made in integrate-and-fire models. These simple spiking models are derived by reducing Hodgkin-Huxley equations using bifurcation methodologies~\cite{Izhikevich}. Thus, this model can accurately recreate the regular spiking, fast spiking and intrinsically bursting firing patterns displayed by neurons in a rat's motor cortex. Note that the equations to model the neuron in the simple spiking setting may be different if only one neuron is to be simulated~\cite{Izhikevich}.

As motivated above, to simulate individual neurons, we use the model developed in~\cite{Izhikevich},
\begin{align}
\dot v &= 0.04v^{2} + 5v + 140 - u + I, \label{eq:neumodel_v}\\
\dot u &= a(bv - u), \label{eq:neumodel_u}
\end{align}
with after spike resetting,

\begin{align}
    \mbox{if}\quad v \geq 30\quad \mbox{then}\quad \begin{cases}
                v\rightarrow c\\
                u\rightarrow u+d.
             \end{cases}
             \label{eq:reset}
\end{align}
Here $v$ is the membrane potential of an individual neuron, $u$ is the membrane recovery variable that provides negative feedback to $v$~\cite{Izhikevich} as is evident from Eqn.~\ref{eq:neumodel_v}. In Eqn.~\ref{eq:neumodel_u}, $a$ (typical value of $a=0.02$) sets the time scale of the recovery of $u$ and $b$ quantifies the coupling between $u$ and $v$ (typical value $b=0.2$). Greater the value of $b$ the stronger the coupling between the variables. The synaptic currents are represented through variable $I$.

The neuron reset rule is quantified in Eqn.~\ref{eq:reset}. In particular, when $v\geq 30$, $v$ is reset to a value of $c$ (typical value $c=-65$ mV). Parameter $d$ in Eqn.~\ref{eq:reset} quantifies the after spike reset value of $u$ (typical value $d=2$). For a more detailed explanation of the model and parameters, we point the reader to~\cite{Izhikevich}. Note that, we use the above model for demonstration purposes. One could easily replace this neural model by more accurate ones such as~\cite{Hodgkin-Huxley}.

Just as in~\cite{Izhikevich}, we achieve heterogeneity by assuming randomness in the parameters for the excitatory and inhibitory neurons in the network. We assume the same randomness as~\cite{Izhikevich}. For excitatory neurons, we assume, $(a_{i},b_{i}) = (0.02,0.2)$ and $(c_{i},d_{i}) = (-65.0,8.0) + (15.0,-6.0)r_{i}^{2}$, where $r_{i}$ is a random variable picked from the uniform distribution on $[0,1]$. For inhibitory neurons, we assume, $(a_{i},b_{i}) = (0.02,0.25) + (0.08,-0.05)r_{i}$ and $(c_{i},d_{i}) = (-65.0,2.0)$. Further, the synaptic connections are represented by the $S$ matrix. When neuron $j$ fires, the value of $v_{i}$ instantaneously changes by the entry $s_{ij}$~\cite{Izhikevich}. We now simulate $800$ excitatory and $200$ inhibitory neurons that are interconnected. To implement this simulation on the cloud, we first construct a MapReduce algorithm to evolve this network of neurons.

\SetKwFor{method}{method}{}{\vspace*{-0.5cm}}
\SetKwFor{class}{class}{}{\vspace*{-0.5cm}}
\setlength{\algomargin}{1em}
%

\section{MapReduce Implementation}
\begin{figure}[htb]
\begin{center}
\includegraphics[width=6in]{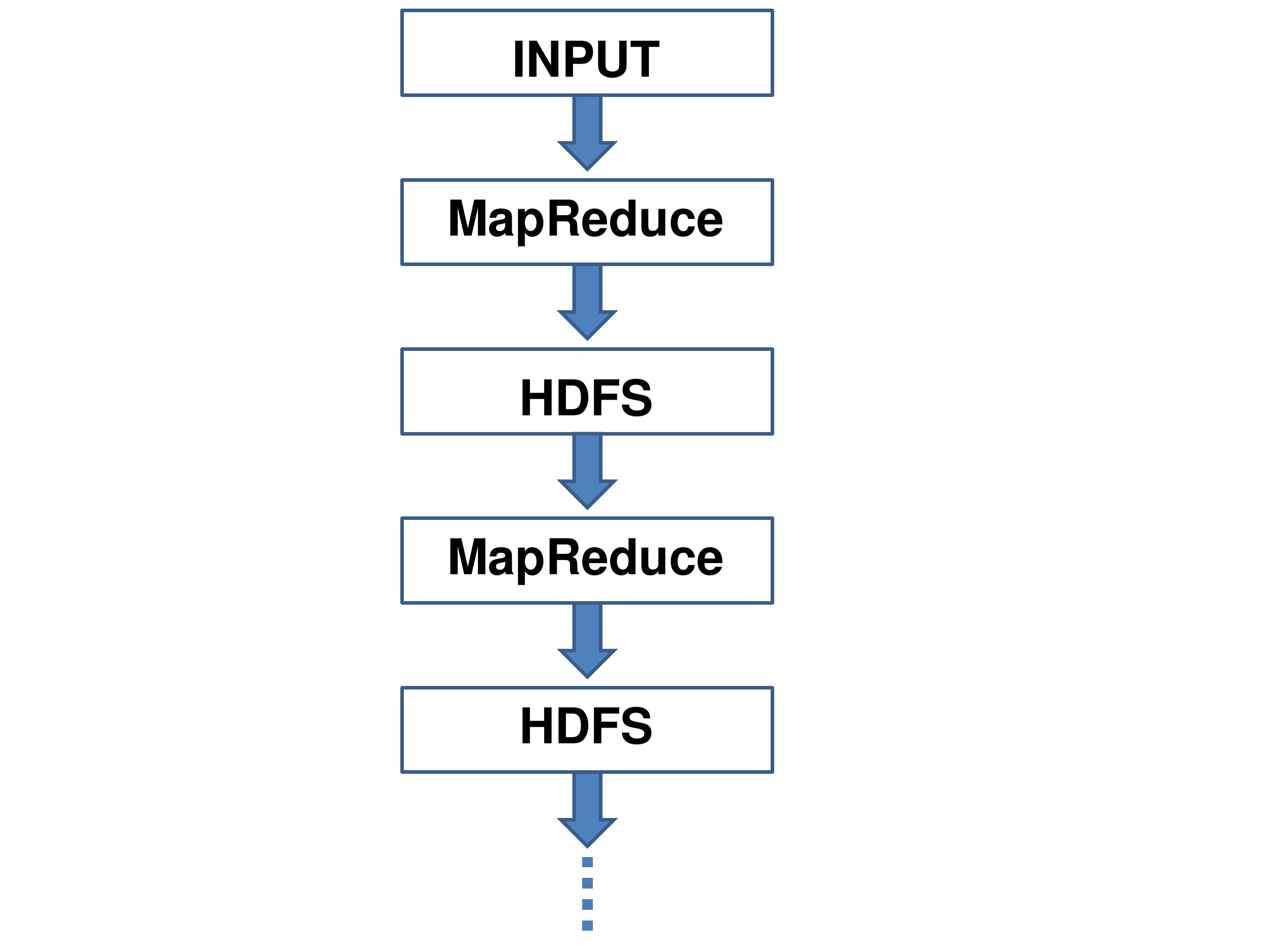}
\caption{Our approach to chaining Hadoop jobs. This will be replaced by other Hadoop job chaining solutions in subsequent implementations.}
\label{Fig:chain}
\end{center}
\end{figure}

To construct a MapReduce approach for the neural modeling of the brain on the cloud, we break the simulation task into the previously described Map and Reduce steps. Each MapReduce step evolves the network of neurons by $1$ ms. To simulate the network for longer time periods, multiple MapReduce steps can be chained together as shown in Fig.~\ref{Fig:chain}. The output of each step is written to HDFS and is read back for the next iteration. Note that, in future work, we are looking to replace the high latency HDFS writing with solutions from the Apache Oozie and Apache Giraph projects.

The approach for a single MapReduce step is illustrated in algorithms~\ref{Algo:Map} and~\ref{Algo:Reduce}. In the map step, the data for each neuron is first read into a data structure for each neuron.  Here the key is the neuron number and the data structure is the corresponding value. Using randomly generated synaptic current values $N.I$, the membrane potential $N.v$ and membrane recovery variable $N.u$ are updated for the current iteration. If the membrane potential $N.v$ exceeds the threshold of $30$ mV, the membrane potential and recovery variables are reset according to Eqn.~\ref{eq:reset}. If the $n$-th neuron has fired, it emits $N.S(j)$ to the $j$-th neuron.

\begin{algorithm}
\DontPrintSemicolon

\class{\textsc{Mapper}}{ \method{\textsc{Map}(Neuron id $n$, Neuron Data
$N$)}{
            $N=$ \textsf{ReadData()};
           \tcp*[h]{Read the data}\;
            \eIf(\tcp*[f]{If Neuron is Excitatory}){$N.TYPE==1$}{
             $N.I = 5\times\textsf{Random}(\left[0,1\right])$ \;
            }
           (\tcp*[f]{If Neuron is Inhibitory}){
            $N.I = 2\times\textsf{Random}(\left[0,1\right])$ \;
           }
            $N.I = N.I + N.Sum$
            $N.v = N.v+0.5(0.04N.v^2+5N.v+140-N.u+N.I)$ \;
            $N.v = N.v+0.5(0.04N.v^2+5N.v+140-N.u+N.I)$ \;
            $N.u=N.u+N.a(N.b v-u)$\;
            $N.Sum \leftarrow 0$\;
            $N.iter\leftarrow N.iter+1$\;
      \If(\tcp*[f]{Check if Neuron has fired}){$N.v\geq 30$}{
             \ForAll{\mbox{Neuron id} m $\in N.S>0$}{
               \textsf{EMIT}(Neuron id $m$, Reset data $N.S(m)$)\tcp*[f]{Emit data}\;
             }
             $N.v \leftarrow N.c$\;
             $N.u \leftarrow N.u + N.d$\;
            }
        \textsf{EMIT}(Neuron id $n$, Neuron Data $N$)\tcp*[f]{Pass Neuron data}\;
    }
} \caption{Mapper for brain simulation on the cloud. The map step for each neuron emits the entire data neuronal data such as $S$, $v$, $u$, $I $ along with the voltage reset $N.S(m)$ that contains firing information.\label{Algo:Map}}
\end{algorithm}

\begin{algorithm}
\DontPrintSemicolon \class{\textsc{Reducer}}{
\method{\textsc{Reduce}(Neuron id m, Neuron Data List $\left[\textsf{p}_{1},\textsf{p}_{2},\hdots\right]$)}{
$M\leftarrow\Phi$\;
\ForAll{$\textsf{p}\in \left[\textsf{p}_{1}, \textsf{p}_{2},\hdots\right]$}{
   \eIf(\tcp*[f]{Recover Data Structure}){IsNode(\textsf{p})}{
     $M\leftarrow \textsf{p}$\;
   }
    (\tcp*[f]{Get firing data}){
     \textsf{p.Sum}$\leftarrow \textsf{p.Sum} + \textsf{p}$\;
    }
     }
  }
} \caption{Reducer for the brain simulation on the cloud.
The Reducer step sums the value for $N.S$ over all the NodeData.\label{Algo:Reduce}}
\end{algorithm}

\begin{figure}[htb]
\begin{center}
\includegraphics[width=6in]{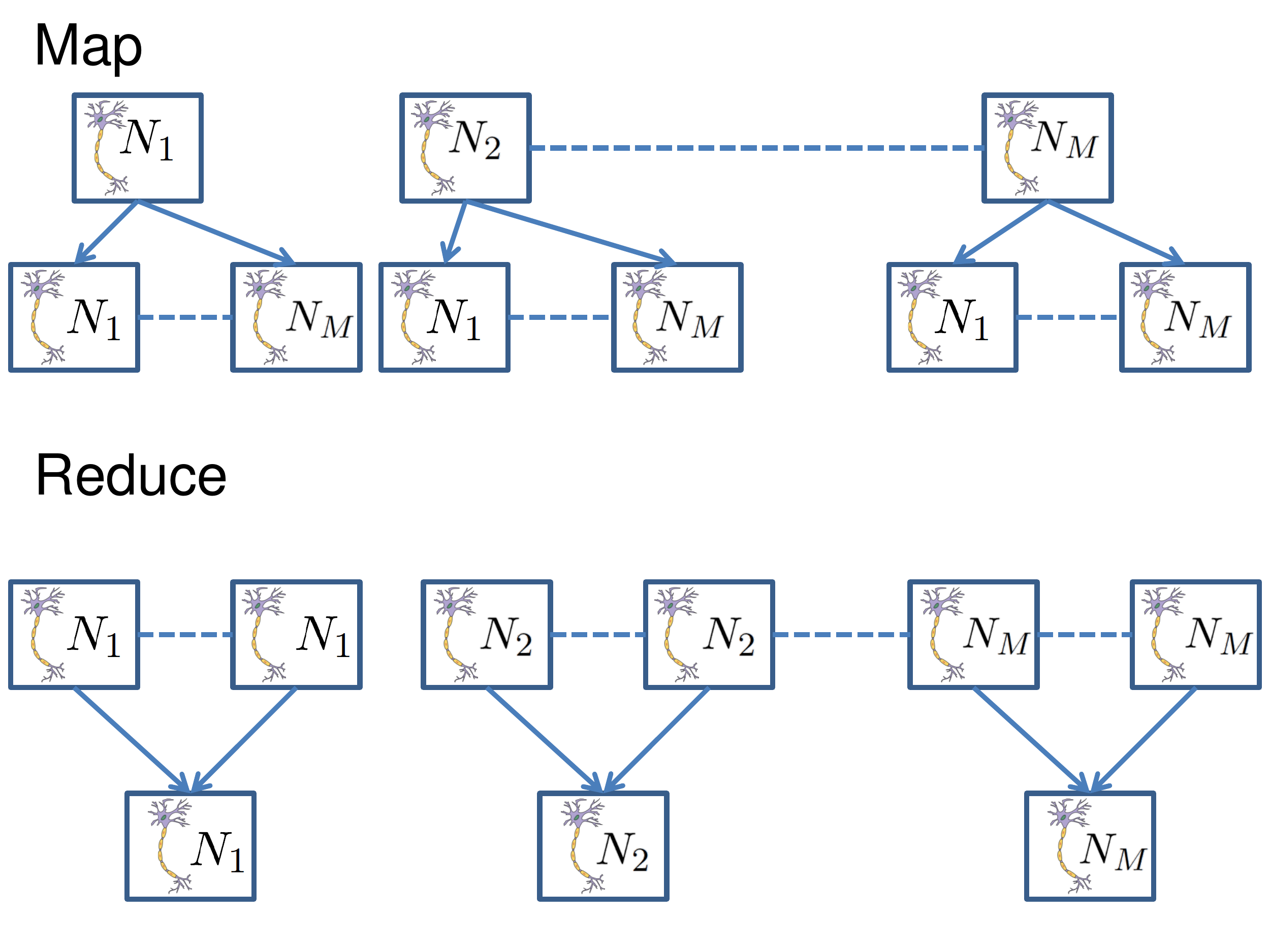}
\caption{MapReduce algorithm for simulating interconnected neurons on the cloud.}
\label{Fig:MRBrain}
\end{center}
\end{figure}

The second step in the process is the reduce step. The reducers collect the outputs of all the neurons and compute the sum of the input synaptic current $N.Sum$ for each neuron. Note that $N.S$ contains the connections of the neuron number $n$ to the rest of the neurons in the network~\cite{Izhikevich}. The algorithmic approach for simulating the network of interconnected neurons in MapReduce is shown in Fig.~\ref{Fig:MRBrain}.

\section{Results} 
\begin{figure}[htb]
\begin{center}
\includegraphics[scale=0.6]{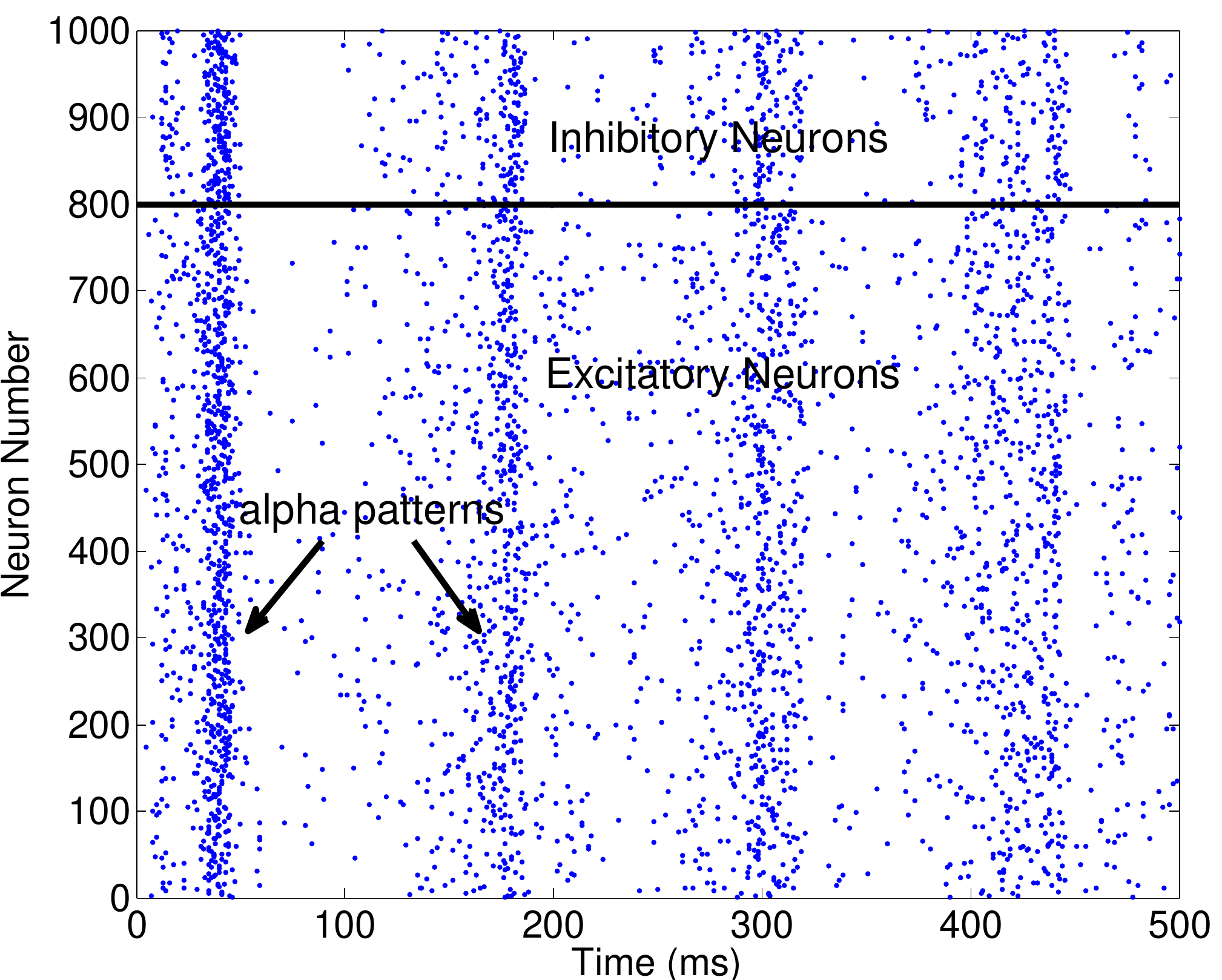}
\caption{The firing in a network of $1000$ neurons over a span of $500$ ms. Computation performed using Hadoop.}
\label{Fig:firing}
\end{center}
\end{figure}

\begin{figure}[htb]
\centering
\includegraphics[width=5in]{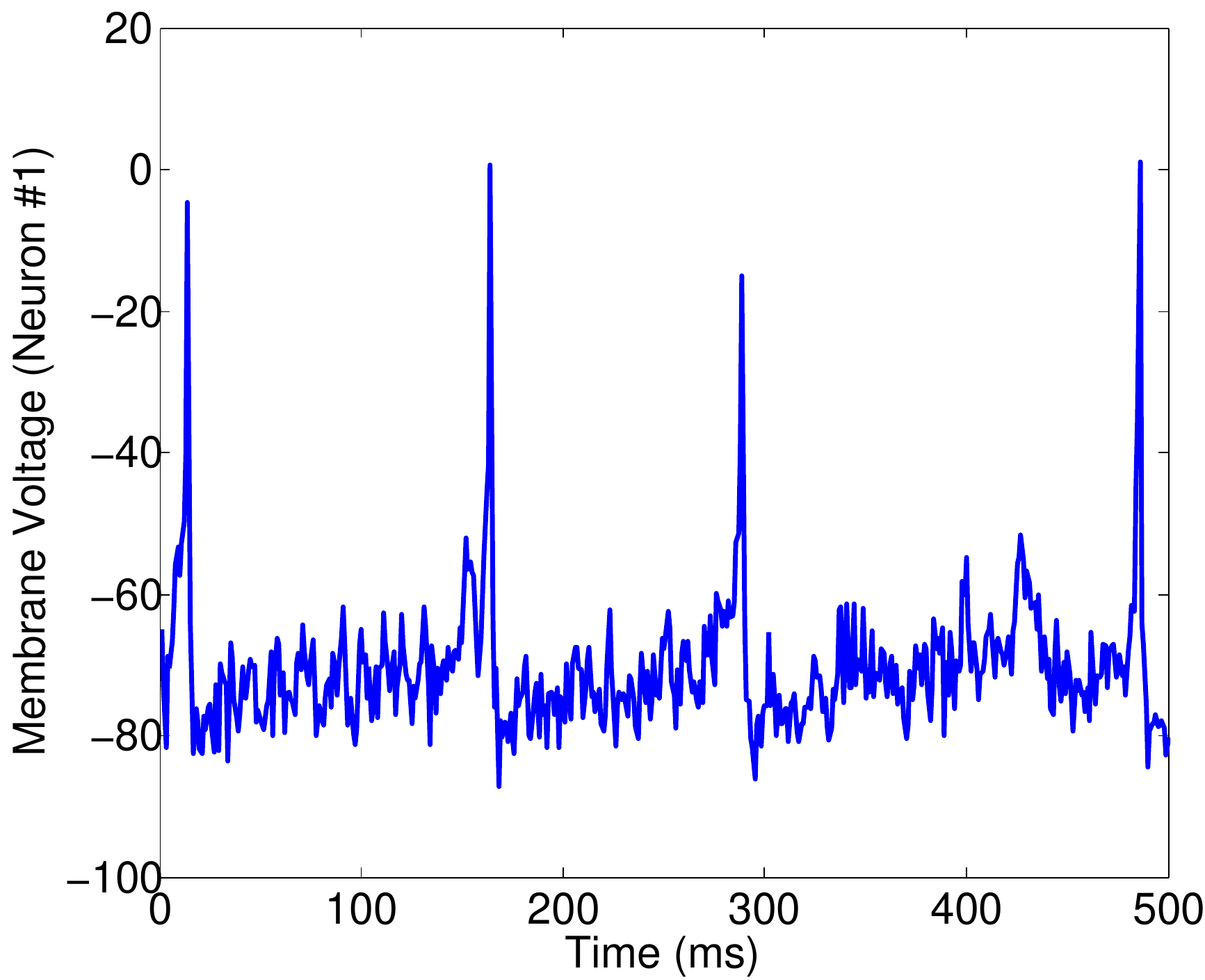}
\caption{The membrane voltage $v$ of neuron number $1$ over a span of $500$ ms. Computation performed using Hadoop.}
\label{Fig:volt1}
 \end{figure}

\begin{figure}[htb]
\centering
\includegraphics[width=5in]{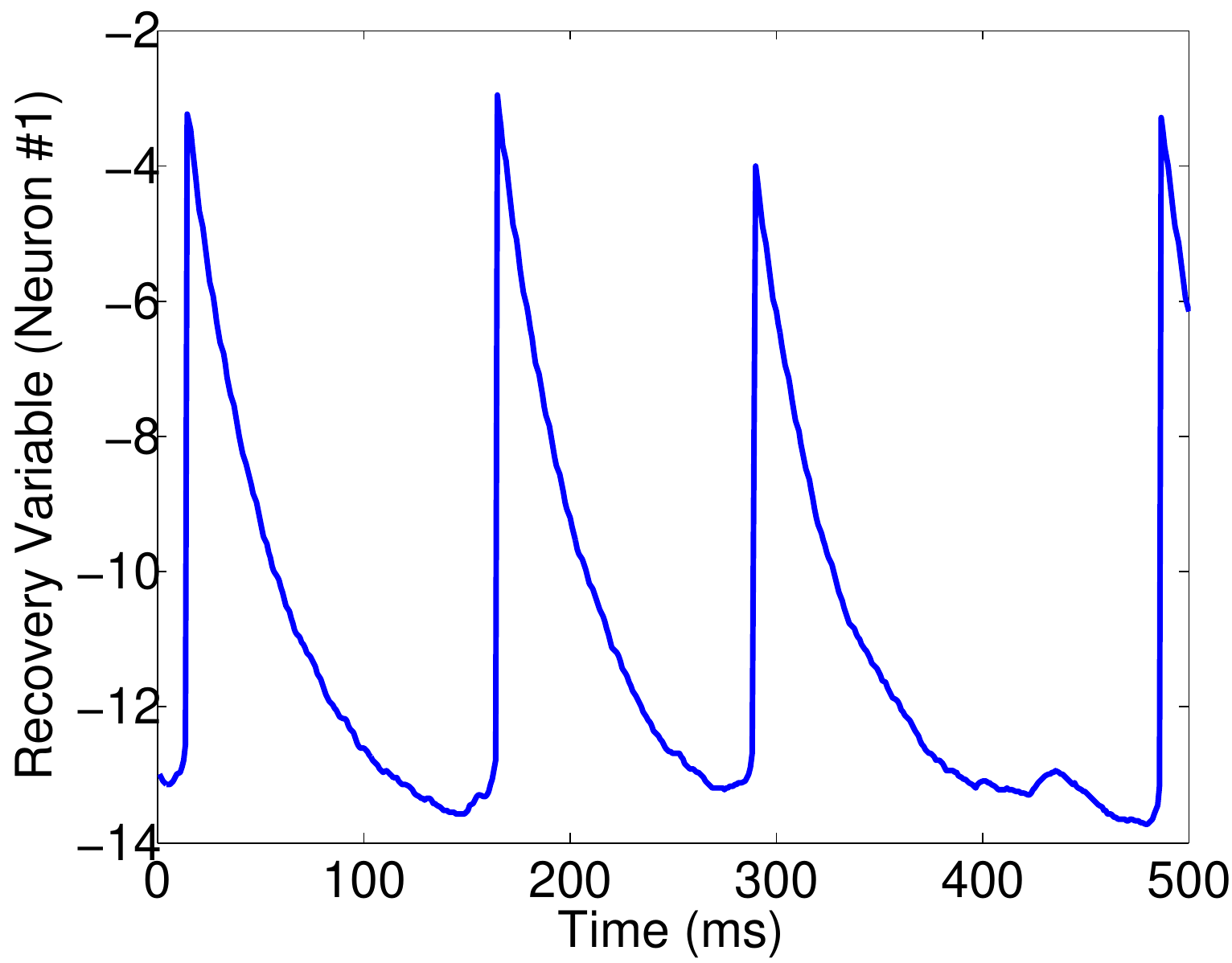}
\caption{The recovery variable $u$ of neuron number $1$ over a span of $500$ ms. Computation performed using Hadoop.}
\label{Fig:recovery1}
 \end{figure}

We now discuss the results from running the Hadoop code for the network of $1000$ neurons. We run the code for $500$ ms and obtain statistically identical results as in~\cite{Izhikevich}. The model is able to capture the synchronization of the neurons. In particular, Fig.~\ref{Fig:firing} shows the synchronous firing of the neurons in the alpha frequency range ($10$ Hz). Typical (using neuron number $1$ as an example) evolution  of the membrane voltage $v$ and recovery variable $u$ is shown in Figs.~\ref{Fig:volt1} and~\ref{Fig:recovery1} respectively.
\hspace{-3in}

The computation was done on a single processor machine with a virtualized cloud (single node cluster) for testing. As expected, the computation time was slower (by a factor of $5$) than the regular code since both the size of the problem and the computing resources were significantly smaller than what would make scalability evident. However, for larger problems (millions of neurons) on the cloud, the approach is expected to provide significant speedup over monolithic implementations. This work demonstrates the feasibility of using the cloud infrastructure for scientific computing. Note, however, we do not claim that the results on the cloud will be comparable to high performance clusters for scientific computing~\cite{PerformanceHPC}. However, we do envision that the costs will be significantly lower if one were to use commodity hardware. Moreover, these costs are expected to reduce over time as cloud computing is embraced and becomes more mainstream.

\section{Conclusions and Future Work}
Cloud computing is becoming increasingly popular for multiple solutions in the IT world. Over the last couple of years, multiple cloud service providers such as Amazon, Google and Microsoft have entered the market with various solutions. Traditionally, this technology has been used to host user data and perform basic analysis. However, this paradigm is expected to change over the next few years. The cloud is expected to not only host data, but also perform computations that can be used for inference and prediction. As cloud technologies are improved, they may have an unexpected application in traditional scientific computation.

In this work, we used the MapReduce paradigm to simulate a network of $1000$ ($800$ excitatory and $200$ inhibitory) neurons. We used the simple spiking model from~\cite{Izhikevich} to model the dynamics of a single neuron. In the future, we intend to use Hodgkin-Huxley equations~\cite{Hodgkin-Huxley} to perform such computations. In particular, we use Hadoop (version $0.20.2$) to perform the computations. The Map step is used to update the equations for individual neurons and the Reduce step is used to aggregate the results. As expected, we find that the results obtained from the MapReduce implementation are accurate.

Our primary computational overhead is the reading and writing to HDFS that needs to be performed at each step. Our current work is focused on replacing this step using graph computation solutions such has Giraph, Pregel and Oozie. We are also identifying other scientific computing problems that can be solved on the cloud. Additionally, Yarn, the next generation implementation of MapReduce in Hadoop, is also expected to provide significant improvements in computational time. Thus, we expect the computational and financial cost of using the cloud to reduce over time. Not only will this make the cloud an increasingly attractive solution for scientific research, but also may provide cheap computing infrastructure to institutions in the third world.

\bibliographystyle{unsrt}
\bibliography{Brain_MapReduce}
\end{document}